\documentclass[aps,prb,reprint,superscriptaddress,showpacs,citeautoscript]{revtex4-2}

\usepackage[dvipdfmx]{graphicx}
\usepackage{color}

\usepackage{amsmath}
\usepackage{accents}
\usepackage{bm}

\makeatletter
\def\widebar{\accentset{{\cc@style\underline{\mskip10mu}}}}
\makeatother

\begin{document}

\title{Degenerate Soft Modes and Selective Condensation in BaAl$_2$O$_4$ via Inelastic X-ray Scattering}


\author{Yui~Ishii}
\email{yishii@mat.shimane-u.ac.jp}
\affiliation{Co-Creation Institute for Advanced Materials, Shimane University, Shimane 690-8504, Japan}

\author{Arisa~Yamamoto}
\affiliation{Department of Materials Science, Osaka Metropolitan University, Osaka 599-8531, Japan}

\author{Alfred Q. R.~Baron}
\affiliation{Materials Dynamics Laboratory, RIKEN SPring-8 Center, Sayo, Hyogo 679-5148, Japan}
\affiliation{Precision Spectroscopy Division, SPring-8/JASRI, Sayo, Hyogo 679-5198, Japan}

\author{Hiroshi~Uchiyama}
\affiliation{SPring-8/JASRI, Sayo, Hyogo 679-5198, Japan}

\author{Naoki~Sato}
\affiliation{Research Center for Materials Nanoarchitectonics (MANA), National Instititue for Materials Science (NIMS), Tsukuba, Ibaraki 305-0044, Japan}


\date{\today}

\begin{abstract}

BaAl$_2$O$_4$ is a ferroelectric material that exhibits structural quantum criticality through chemical composition tuning. 
Although theoretical calculations and several diffraction experiments have suggested the involvement of a soft mode in its ferroelectric structural phase transition, direct experimental verification is still lacking.
In this study, we successfully observed two soft modes of BaAl$_2$O$_4$ using x-ray inelastic scattering, providing direct experimental evidence for their role in the structural phase transition.
Furthermore, we reveal that the soft modes at the M and K points are nearly degenerate in energy, indicating a delicate balance in which either mode could potentially freeze.
The K-point mode simultaneously softens toward the transition temperature ($T_{\rm C}$) in a manner nearly identical to the M-point mode.
However, the phase transition condenses only at the M point, with the M-point mode stabilizing as an acoustic mode in the low-temperature structure and the K-point mode hardening as temperature decreases.

\end{abstract}


\maketitle

\section{Introduction}
	The emergent physical properties associated with the suppression of structural phase transition have recently attracted growing interest, with {\it structural} quantum criticality as the key concept.
	This idea was first explored in the 1970s in the context of quantum paraelectrics such as SrTiO$_3$, KTaO$_3$ and EuTiO$_3$ \cite{STO_quantumpara,KTO_quantumpara,ETO}.
	The pristine perovskite compounds themselves do not exhibit the ferroelectric phase transition, which appears by chemical substitution or applying external stress.
	The dielectric susceptibility exhibits a divergent increase toward zero temperature in the pristine compound due to the suppression of the ferroelectric soft mode to condense.	
	They have consistently drawn attention from both experimental and theoretical perspectives, particularly in relation to the emergence of polar superconductivity in lightly doped states \cite{STO_super1, Ferroelectric_Super1, Kanasugi, Edge}.

	Studies have reported other structural quantum materials such as the intermetallic compound (Sr$_{1-x}$Ca$_x$)$_3$Rh$_4$Sn$_{13}$, which exhibits strong-coupling superconductivity in proximity to the structural quantum critical point (sQCP) \cite{Goh_PRL114,Goh_PRL115,Ishii_jpsj}. 
	Materials with the CdI$_2$-type structure have also been recognized as candidates for exhibiting structural quantum criticality, although their structural phase transitions are of first-order character \cite{IrTe2,AuPdxTe2}.
	More recently, the interplay between the polar structural quantum criticality and thermoelectric performance in MoTe$_2$ has gained increasing attention \cite{MoTe2_SciAdv,MoTe2_PRB}.
	Thus,  structural quantum critical phenomena are now a growing research area in materials science.

	A structural quantum phase transition is a continuous transition that is driven by quantum fluctuation.
	Therefore, for a structural quantum critical point to appear, the associated structural phase transition must be of second-order character.
	Soft-mode condensation is widely known as a typical origin of second-order structural phase transitions.
	Here, we use the term {\it soft mode} to refer a phonon mode whose frequency decreases when approaching a structural phase transition. 
	Among the materials mentioned above, substituted SrTiO$_3$ and KTaO$_3$ exhibit a structural quantum critical point at $T=0$ involving an optical soft mode located just above the acoustic modes \cite{Yamada_JPSJ26, Shirane_KTO}.
	The structural phase transitions reported in the materials with CdI$_2$-type structure are generally attributed to instabilities in electronic system \cite{IrTe2_Pd, IrTe2_calc}.
	However, even in these materials, soft modes may arise from the coupling between electronic instabilities and the lattice degrees of freedom.
	Indeed,  inelastic x-ray scattering has revealed that the CDW transition in TiSe$_2$ is accompanied by phonon softening \cite{TiSe2}.

	Here, we focus on Ba$_{1-x}$Sr$_x$Al$_2$O$_4$, which exhibits a structural quantum critical point \cite{Ishii_PRB106}.
	The parent material BaAl$_2$O$_4$ is an improper ferroelectric with a Curie temperature ($T_{\rm C}$) of 450 K \cite{Stokes}.
	Although it has not been experimentally confirmed, calculations have indicated the presence of two unstable phonon modes at the M and K points of one of the transverse acoustic branches in the high-temperature phase \cite{Ishii_PRB93}.
	The ferroelectric phase transition is believed to occur through improper coupling between the M-point mode and a polar mode at the $\Gamma$ point \cite{Perez-Mato}.
	The crystal structure of the high-temperature phase is characterized by a network structure consisting of corner-sharing AlO$_4$ tetrahedra with hexagonal cavities occupied by Ba ions.
	The space groups of the high-temperature and low-temperature phases are $P6_322$ and $P6_3$, respectively.
	This structural phase transition has been characterized as a second-order transition \cite{Ishii_PRB93,Perez-Mato}.

	The Ba site can be fully substituted with Sr.
	The ferroelectric phase completely disappears with a small amount of Sr substitution \cite{Ishii_SciRep,Kawaguchi_PRB,Ishii_PRB94}, 
	and the sQCP is considered to be located near $x$ = 0.1 in Ba$_{1-x}$Sr$_x$Al$_2$O$_4$.
	A particularly intriguing aspect of this material is the emergence of amorphous-like properties at Sr compositions higher than the sQCP composition.
	At Sr compositions exceeding the sQCP, an excess lattice specific heat and a plateau in thermal conductivity are observed, which are typical characteristics of amorphous solids \cite{Ishii_PRB106,Ishii_PRM}.
	Synchrotron x-ray pair distribution function analysis has revealed deviations of specific atomic positions from the average structure.
	Inelastic neutron scattering measurements on powder samples have demonstrated that the phonon-related peaks are significantly damped at the compositions exceeding the sQCP, resulting in a spectrum resembling the boson peak commonly observed in amorphous solids \cite{boson1,boson2,boson3}.
	These observations suggest that the phenomenon occurring at the structural quantum critical point may essentially be attributed to the incoherent freezing of the soft mode \cite{Ishii_PRB106}.
	Therefore, experimentally verifying the presence of a soft mode in the parent compound BaAl$_2$O$_4$, and closely tracking its development, is important for elucidating the nature of structural fluctuations that anticipated at the sQCP.
	In this study, we directly observe the soft mode involved the structural phase transition of BaAl$_2$O$_4$ using meV-resolution inelastic x-ray scattering (IXS).

\section{Experimental}
	High-quality BaAl$_2$O$_4$ single crystals were grown using the self-flux method with a platinum crucible. 
	The synthesis procedure is described in detail in the reference \cite{Ishii_PRB93}.
	Inelastic x-ray scattering measurements were conducted at BL35XU \cite{BL35XU-1, BL35XU-2} of SPring-8.
	Single crystals of BaAl$_2$O$_4$ with a size of approximately 200 $\mu$m were mounted at the tip of a quartz glass capillary with silver paste.
	A Si(11 11 11) backscattering geometry was used.
	The incident x-ray energy was set at 21.747 keV with an energy resolution of 1.5 meV full width at half maximum.
	Based on the analysis of phonon eigenvectors described later, data were collected from the (008) $\Gamma$ point to the M and K points, as shown in Fig. \ref{eigen}(a) by arrows, with the scattering vectors $Q = (h \; 0 \; 8)$ and $(h \; h \; 8)$, respectively.
	The typical $Q$ resolutions were (0.05 0.02 0.10) and (0.03 0.03 0.10) in ($hkl$), respectively. 
	The temperature was varied between room temperature and 650 K.
	To fit the obtained spectra, we employed a damped harmonic oscillator (DHO) model, as is described in detail in Supplemental Material \cite{Supple} (including references \cite{Baron2019,James1975,baron_absfit}).
	
	Calculations were performed using density functional theory (DFT) as implemented in the Vienna ab initio simulation package (VASP) \cite{vasp,vasp1,vasp2,vasp3}, and phonon properties were subsequently evaluated with Phonopy \cite{phonopy,phonopy2,phonopy3,phonopy4}.
	Projector augmented wave pseudopotentials \cite{pseudopotential1,pseudopotential2} and the generalized gradient approximation functional with the Perdew-Burke-Ernzerhof parametrization revised for solids (GGA-PBEsol) \cite{GGA} were chosen for the DFT total energy calculations. 
	The unit cell of BaAl$_2$O$_4$ ($P6_322$) was fully relaxed until the residual forces became less than 10$^{-4}$ eV/\AA. 
	A $k$-point grid of $7\times7\times4$ and an energy cutoff of 500 eV for the primitive cell were used for the structure relaxation. 
	To extract the second interatomic force constants (IFCs), finite difference method with as $2\times2\times2$ supercell containing 112 atoms based on the fully relaxed structure to create displacement-force datasets. 
	The IFCs and phonon dispersion were calculated using Phonopy. 
	A non-analytical term correction of optical phonon frequencies derived from long-range interaction \cite{non-Analy} was taken into account.
	The lattice vectors chosen in the above calculation are defined in Supplemental Material \cite{Supple}.

\section{Results and Discussion}
\begin{figure}[t]
\begin{center}
\includegraphics[width=85mm]{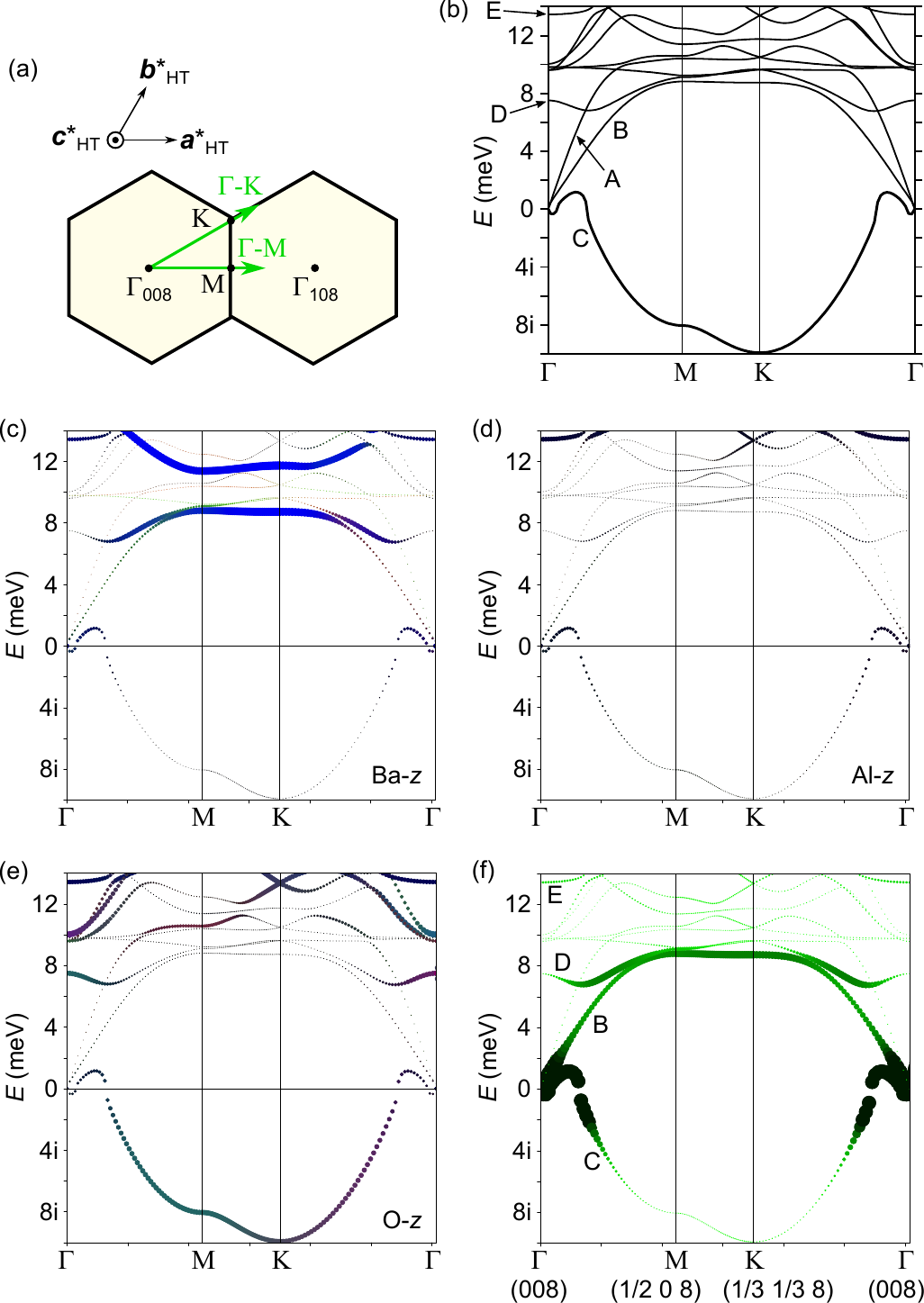}
\caption{\label{eigen} 
(a) ($hk$8) cross section of the reciprocal space of the hexagonal lattice.
Inelastic x-ray scattering experiments were performed along the $\Gamma$--M and $\Gamma$--K directions, with the scattering vectors $\bm{Q} = (h\; 0\; 8)$ and $\bm{Q} = (h\; h\; 8)$, respectively, as shown by green arrows.
(b) Calculated phonon dispersion of the high-temperature phase of  BaAl$_2$O$_4$ obtained along the $\Gamma$--M--K--$\Gamma$ path. The three acoustic branches are labeled A, B, and C, and the optical branch just above them is labeled D. 
Branch E is considered to be an optical branch corresponding to the weak scattering observed in the spectrum at small $h$ around $\pm$13 meV in Fig. 2.
Panels (c), (d), and (e) visualize the magnitudes of the phonon eigenvectors of Ba, Al, and O atoms in the $z$ direction, respectively (see also the main text).
(f) Dynamical structure factor $S(\bm{Q}, E)$ calculated along the $\Gamma$ (008) -- M (1/2 0 8) -- K (1/3 1/3 8) -- $\Gamma$ (008) path.
Larger marker sizes and darker colors indicate higher values of $S(\bm{Q}, E)$.
}
\end{center}
\end{figure}

	Figure \ref{eigen}(b) displays the calculated phonon dispersion of the high-temperature phase of BaAl$_2$O$_4$.
	Several branches in the figure are labeled A through E.
	The acoustic branch C shows a dynamical instability, as plotted by an imaginary frequency, which is broadly consistent with the previous report \cite{Ishii_PRB93}.
	Fig. \ref{eigen}(c) highlights the magnitudes of the $z$-components of the phonon eigenvectors for Ba atoms. 
	Here, the marker size represents the magnitude of the $z$-component at each $\bm{k}$-vector, with larger markers indicating higher values.
	These values are displayed as the sum over all Ba atoms within the unit cell.
	In addition, the RGB color coding serves as a complementary visualization of the relative contributions of $x$, $y$, and $z$ atomic displacements to the phonon modes; red, green, and blue intensities correspond to the $x$, $y$, and $z$-component magnitudes, respectively.
	In the same manner, Figs. \ref{eigen}(d) and \ref{eigen}(e) present the magnitudes of the $z$-components of the phonon eigenvectors for Al and O atoms, respectively.
	The calculated phonon eigenvectors of Ba, Al, and O atoms resolved into their $x$-, $y$-, and $z$-components are summarized in Figs. S1--S3 \cite{Supple}.
	For example, Figs. S1(a) and S1(b) highlight the $x$- and $y$-components, respectively, of the phonon eigenvectors for Ba atoms.
	From this analysis, the branch labeled C, which shows signs of instability, is determined to mainly possess the $z$-polarized characteristic across the $k$-range of interest, while the $x$ and $y$ components primarily determine the nature of the branches labeled A and B, respectively.

\begin{figure}[t]
\begin{center}
\includegraphics[width=85mm]{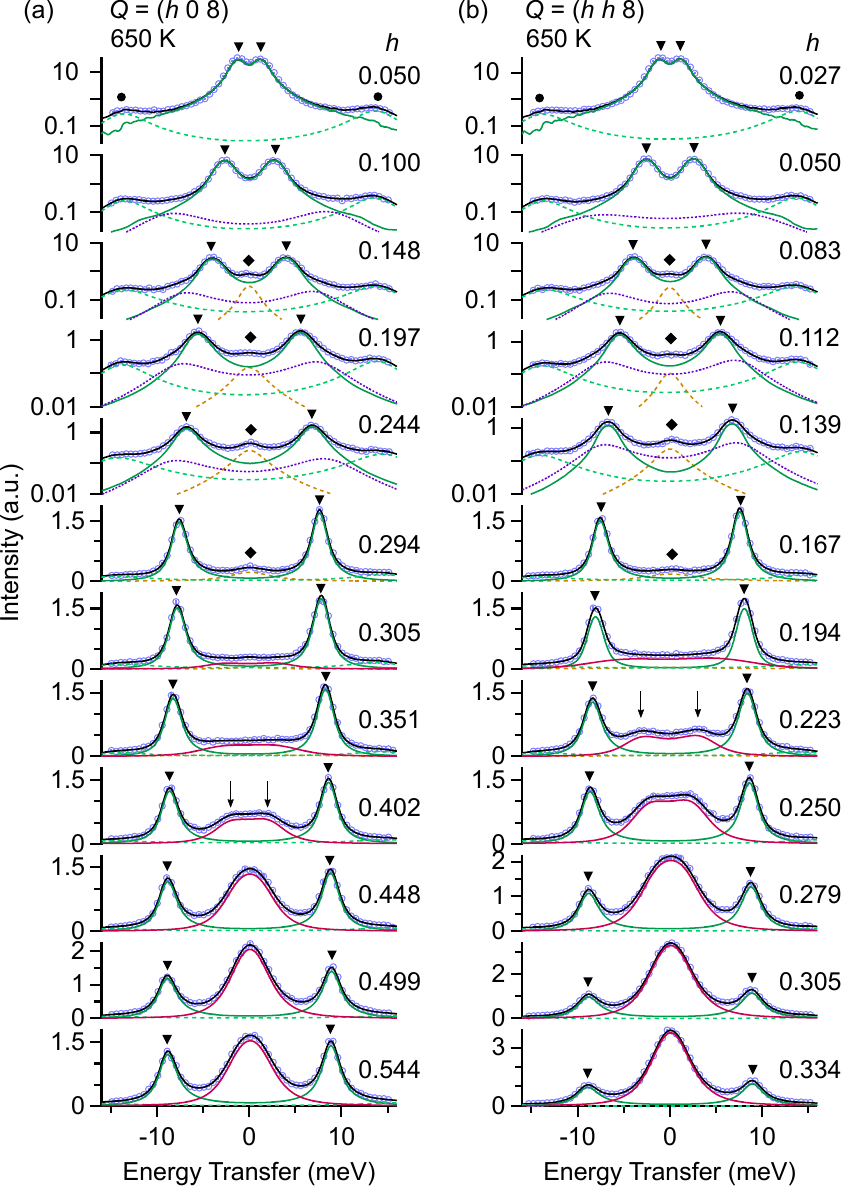}
\caption{\label{spectra} 
(a) IXS spectra measured at 650 K with various scattering vectors $\bm{Q} = (h\; 0\; 8)$ along the $\Gamma$-M direction.
The inelastic peaks indicated by triangles are attributed to branch B shown in Fig. 1(f).
Additionally, small inelastic peaks are observed around $\pm$13 meV, as indicated by circles, which is attributed to branch E.
The inelastic peaks marked by arrows in the spectrum at $h$ = 0.402 correspond to the M-point soft mode.
(b) IXS spectra measured at 650 K with various scattering vectors $\bm{Q} = (h\; h\; 8)$ along the $\Gamma$-K direction.
Similar to the spectra observed in Panel (a), inelastic peaks are observed as indicated by triangles and circles, which are attributed to branch B and E, respectively.  
The inelastic peaks indicated by arrows in the spectrum at $h = 0.223$ correspond to the K-point soft mode.
}
\end{center}
\end{figure}

	Fig. \ref{eigen}(f) presents the simulation of dynamical structure factor $S(\bm{Q}, E)$ calculated along the path of $\Gamma$(008)--$M$(1/2 0 8)--$K$(1/3 1/3 8)--$\Gamma$(008). 
	This calculation includes only the contribution of one-phonon process \cite{AM}.
	The magnitude of $S(\bm{Q}, E)$ is represented by the size of markers.
	The calculated values of $S(\bm{Q}, E)$ at $T=0$ for acoustic phonons near the $\Gamma$ point are saturated at a maximum threshold due to extremely large values around the $\Gamma$ point.		
	Although the dominant vibrational component of the branch B near the $\Gamma$ point is in the $y$-direction, as shown in Fig. S2(b), the dynamical structure factor $S(\bm{Q}, E)$ of the branch B is also relatively large, as shown in Fig. \ref{eigen}(f).
	This reflects the fact that the branch B contains a certain degree of $z$-component, as shown in Figs. \ref{eigen}(c)--\ref{eigen}(e).

\begin{figure*}[t]
\begin{center}
\includegraphics[width=175mm]{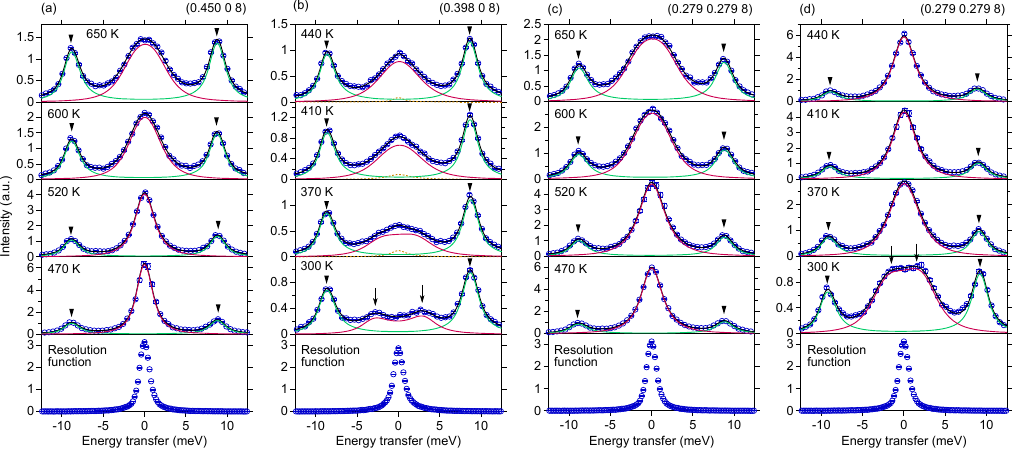}
\caption{\label{softmode} 
Temperature dependence of inelastic scattering spectra measured near the M and K points.
Panels (a) and (c) show the spectra measured above $T_{\rm C}$, while Panels (b) and (d) show those measured below $T_{\rm C}$.
The scattering vectors are: (a) $\bm{Q}$ = (0.45\;0\;8), (b) $\bm{Q}$ = (0.398\;0\;8), (c) and (d) $\bm{Q}$ = (0.279\;0.279\;8).
The solid black curves represent the fitting results using the DHO model.
Resolution functions corresponding to each spectrum are shown at the bottom of each panel. 
The parameters of peak center and peak width obtained from the fit are included in Supplemental Material \cite{Supple}.
}
\end{center}
\end{figure*}

	Figure \ref{spectra}(a) shows the measured IXS spectra at each ${\bm Q}$ along the $\Gamma$-M direction at 650 K.
	The spectra from $h=0.050$ to 0.244 are shown with a logarithmic scale on the vertical axis. 
	The observed peaks were assigned based on the prediction of $S(\bm{Q},E)$ shown in Fig. \ref{eigen}(f).
	The spectra were successfully resolved into several peaks using the DHO model as shown in Fig. \ref{spectra}, where the phonon contributions from branches B, D, and E are represented by solid green, dotted purple, and broken green curves, respectively. 
	Overall, the fitting results reproduce the experimental spectra well.
	The spectra at $h=$ 0.148--0.294 include a small elastic peak at $E=0$, as shown by diamonds.
	Because the crystal used is quite small, about 200 $\mu$m in diameter, we believe that these peaks originate from scattering from Ag paste used to mount the crystal.
	This contribution is also successfully separated, as shown by a broken orange curve. 
	The inelastic peaks from branch B are marked by triangles.
	The center of the Stokes peak coming from branch B shifts toward higher energies as $h$ increases, following its dispersion relation shown in Fig. \ref{eigen}(f).
	The small inelastic peaks coming from branch E, observed approximately at $E=\pm 13$ meV, are indicated by circles.

	The calculated $S(\bm{Q}, E)$ shown in Fig. \ref{eigen}(f) predicts that the optical branch D also produces strong scattering at the $\bm{Q}$ vectors along the $\Gamma$-M direction, resulting in spectral overlap with the inelastic peak from the acoustic branch B for $h=$ 0.100--0.244.
	Indeed, for these spectra, good fits required including the contribution of branch D.
	However, due to the significant overlap of the peak of branch D to the peak of branch B, it was challenging to reliably determine the center of the peak of branch D.

	At $h=0.402$, as indicated by arrows, additional inelastic peaks are observed in the low-energy region on either side of $E=0$, partially overlapping with each other.
	With increasing $h$, the low-energy peaks merge into a single broad peak, while at slightly lower $h$ values, $h =0.305$ and 0.351, weak scattering features are also observed in the same low-energy region.
	These features are likely attributable to inelastic scattering originating from the same phonon, suggesting a continuous evolution of the phonon with increasing $h$.
	Fig. \ref{spectra}(b) displays the IXS spectra acquired at 650 K at each $\bm{Q}$ along the $\Gamma$-K direction.
	These spectra exhibit behavior similar to that observed along the $\Gamma$-M direction.
	The peaks marked with triangles are attributed to the acoustic phonon of branch B.
	For $h \geq 0.223$, the spectra exhibit additional inelastic peaks in the low-energy region, as marked by arrows.
	Assuming that these peaks originate from the soft mode, the overlapped low-energy peaks were resolved into two distinct inelastic peaks by using a DHO model.
	The obtained fit results for the peaks are shown by solid red curves.
	
	Here, it should be noted that the experiment was performed in a configuration emphasizing the $z$-displacement.
	Therefore, the acoustic phonons of branch C, which mainly has motions in the $z$ direction, should be observed near $\Gamma$.
	However, only the peaks indicated by downward triangles in Fig. \ref{spectra} are observed and are assigned to branch B based on the $S(Q, E)$ calculation shown in Fig. 1(f).
	As discussed later, the peaks indicated by downward triangles at $h<0.25$ ($\Gamma$-M) and $h<0.167$ ($\Gamma$-K) are considered to also include contributions from branch C that nearly overlap with branch B.

	To gain further insight into the behavior of the low-energy peaks, we tracked their temperature dependence and performed a detailed analysis of the peak positions.
	Fig. \ref{softmode}(a) presents the temperature-dependent spectra measured at $\bm{Q}$ = (0.450\;0\;8) on the $\Gamma$-M line.
	The panels represent the spectra obtained at several temperatures above the phase transition temperature $T_{\rm C}$.
	The overlapping spectrum was successfully decomposed into two distinct inelastic peaks, which are shown as red and green curves.
	The center of the low-energy peaks is found to locate at 2.9, 2.4, 1,9, and 1.3 meV at 650, 600, 520, and 470 K, respectively, gradually shifting toward $E=0$ as temperature approaches $T_{\rm C}$.

	The spectra measured at temperatures below $T_{\rm C}$ are presented in Fig. \ref{softmode}(b).
	To avoid strong elastic scattering from the superlattice reflection at the M point below $T_{\rm C}$ = 450 K, the measurements were performed at $h=0.398$, slightly offset from the exact M-point position.
	At 440 K, the scattering still centers near $E=0$, however, it becomes broader with decreasing temperature. 
	Eventually at 300 K, it clearly splits into two inelastic peaks, as indicated by arrows.
	These changes observed in Figs. \ref{softmode}(a) and \ref{softmode}(b) reflect the typical phonon softening and hardening behavior, respectively.

\begin{figure}[t]
\begin{center}
\includegraphics[width=85mm]{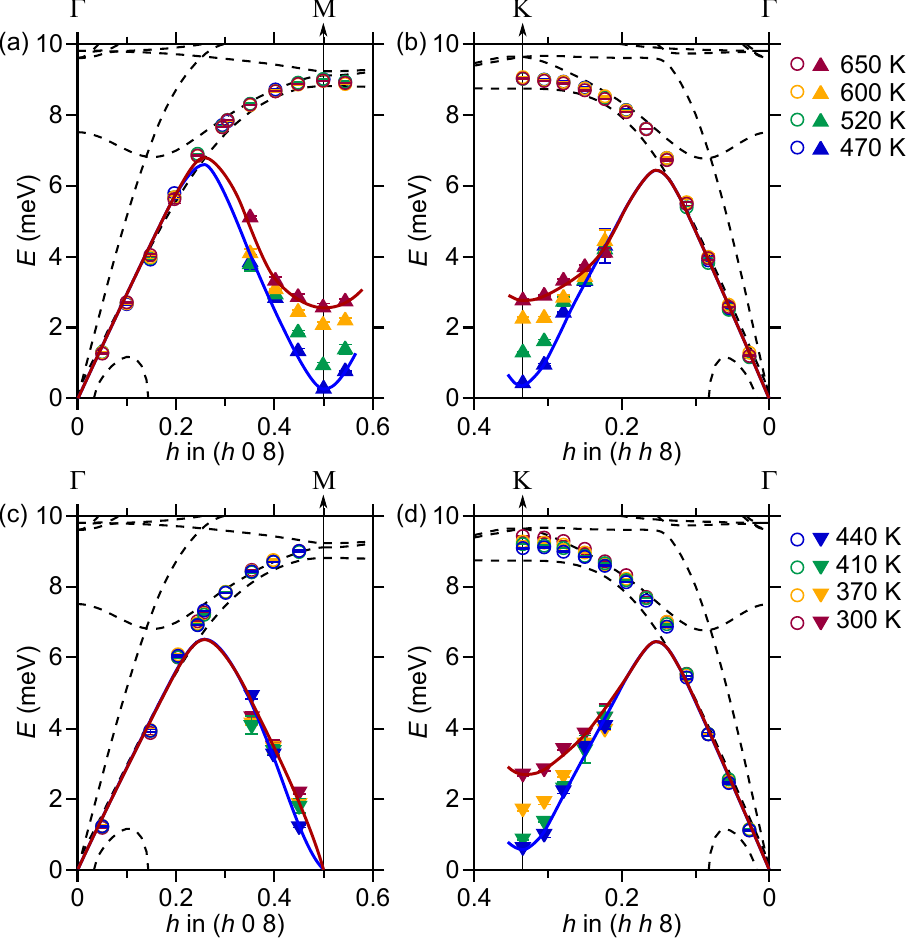}
\caption{\label{disp} 
Experimentally obtained phonon dispersions along the (a, c) $\Gamma$-M and  (b, d) $\Gamma$-K directions. Panels (a) and (b) show the results measured above $T_{\rm C}$, while Panels (c) and (d) show those below $T_{\rm C}$.
The open circles plot the peak positions of the phonon indicated by triangles in Fig. 2 and Fig. 3.
The upward and downward triangles represent the peak positions of the low-energy phonons indicated by arrows in Fig. 2 and Fig. 3.
The broken curves represent the calculated phonon dispersion.
Solid curves are guides to the eye, 
representing the dispersion relation of branch C at representative temperatures estimated as discussion in the main text. 
}
\end{center}
\end{figure}

	Interestingly, similar soft-mode behavior is also observed along the $\Gamma$-K direction.
	IXS spectra at $\bm{Q}$ = (0.279\;0.279\;8) measured above and below $T_{\rm C}$ are shown in Figs. \ref{softmode}(c) and \ref{softmode}(d), respectively. 
	In the same manner as that performed for Figs. \ref{softmode}(a) and \ref{softmode}(b), the spectra were resolved into two inelastic peaks.
	In Fig. \ref{softmode}(c), the center of the low-energy peaks is determined to be at 3.3, 2.8, 2.7, and 2.4 meV at 650, 600, 520, and 470 K, respectively, gradually shifting toward $E=0$ as temperature decreases.
	This fact indicates that, in addition to the phonon softening at the M point, the phonon at the K point also softens toward $T_{\rm C}$.
	Below $T_{\rm C}$, the scattering that merges near $E=0$ gradually broadens as the temperature decreases, and eventually splits into two distinguishable inelastic peaks at 300 K, as shown by arrows.
	All spectra measured at various $\bm{Q}$ vectors along the $\Gamma$-M and $\Gamma$-K directions, along with their temperature dependence, are shown in Supplemental Figs. S7--S10 \cite{Supple}.

	Fig. \ref{disp} shows the values of the peak centers obtained from fitting, plotted along the $\Gamma$-M and $\Gamma$-K.
	The broken curves in the figure represent the calculated dispersion from Fig. \ref{eigen}(b).
	Figs. \ref{disp}(a) and \ref{disp}(b) summarize the results above $T_{\rm C}$, while Figs. \ref{disp}(c) and (d) represent the results below $T_{\rm C}$.
	The temperature dependent low-energy phonons are plotted as upward or downward triangles. 
	In Figs. \ref{disp}(a) and \ref{disp}(b), it is clearly observed that the phonon modes exhibit softening with decreasing temperature in both directions.
	On the other hand, in the results below $T_{\rm C}$, we see a marked difference between the two directions; 
	As shown in Fig. 4(c), along the $\Gamma$-M direction, the phonon energy marked by downward triangles strongly depends on $h$ and shows only a weak dependence on temperature.
	In contrast, as shown in Fig. \ref{disp}(d), along the $\Gamma$-K direction, the phonon energy indicated by downward triangles tends to increase overall as the temperature decreases.

	The dispersion presented as circles in Figs. \ref{disp}(a)--\ref{disp}(d), primarily attributed to branch B, remains nearly unchanged with temperature along both the $\Gamma$-M and $\Gamma$-K directions.
	As revealed by the calculated eigenvectors shown in Fig. \ref{eigen}(c) and Supplemental Figs. S2--S4 \cite{Supple}, branch B is predominantly characterized by the //$y$ component, although it also contains a certain fraction of the //$z$ component. 
	Consequently, as shown in the calculated $S(\bm{Q},E)$ (Fig. \ref{eigen}(f)), branch B is expected to exhibit substantial intensity for scattering vectors $\bm{Q}=(h\; 0\; 8)$ and $(h\; h\; 8)$.
	This expectation is consistent with the strong peaks observed experimentally, indicated by downward triangles in Figs. \ref{spectra} and \ref{softmode}.

	Based on phonon calculations for the low-temperature phase shown in Figs. S5 and S6 \cite{Supple}, the acoustic branches labeled B' and C' in Fig. S5 are dominated by //$y$ and//$z$ components, respectively.
	This correspondence suggests that the two acoustic branches B and C in the high-temperature phase evolve into branches B' and C' in the low-temperature phase, respectively.
	These two branches are nearly degenerate in energy along the $\Gamma$-M' and $\Gamma$-K' within the Brillouin zone of the low-temperature phase.
	Experimentally, the dispersion around the $\Gamma$ point shows no temperature variation above and below $T_{\rm C}$.
	This fact indicates that the dispersion does not exhibit any softening within the observed temperature range.
	Therefore, branch C is expected to have nearly the same energies as branch B around the $\Gamma$ point without showing temperature dependence.
	This implies that, for $h$ values near the $\Gamma$ point in Figs. \ref{spectra} and \ref{softmode}, scattering from branch C overlaps with that from branch B, appearing as the inelastic peaks indicated by downward triangles.

	The dispersion relations of branch C estimated from the above discussion are shown in Fig. 4 for representative temperatures as solid curves.
	According to the previous report \cite{Ishii_PRB93}, single-crystal x-ray diffraction revealed very sharp superlattice reflections at the M point below $T_{\rm C}$. 
	This fact means that the M point in the high-temperature phase becomes a $\Gamma$ point in the low-temperature phase, and the phonon dispersion below $T_{\rm C}$ plotted by triangles reaches $E=0$ at (0.5 0 8).  
	That is, a new acoustic branch emerges from the M point below 440 K, as shown by the solid blue and red curves in Fig. \ref{disp}(c), indicating that the mode responsible for the structural phase transition is exclusively located at the M point.	
	On the other hand, no superlattice reflection is observed at the K point below $T_{\rm C}$ \cite{Ishii_PRB93}.
	This fact means that, although the K-point mode also softens on cooling toward $T_{\rm C}$ as shown in Fig. \ref{disp}(b), it does not freeze; its energy begins to increase again below $T_{\rm C}$, as shown in Fig. \ref{disp}(d).

	As shown in Figs. \ref{disp}(a) and \ref{disp}(b), the two modes exhibit nearly the same energy above $T_{\rm C}$.
	These results immediately raise the question of whether the phonon softening also occurs at all the intermediate $k$ points on the M--K line, not only at the symmetry points, M and K.
	According to our previous x-ray thermal diffuse scattering (TDS) experiments using single crystals \cite{Ishii_PRB93}, the TDS is strongly observed at all $k$-points along the M--K line, and the intensities at these points are temperature-dependent, increasing as the temperature drops toward $T_{\rm C}$. 
	This suggests that the phonons between the M and K points also soften, in addition to the softening observed at the M and K points themselves.
	The phonon calculation shown in Fig. 1(b) also supports this.
	 The simultaneous softening of more than one modes especially along the zone boundary, as observed in the present system, is a rare phenomenon.
	 Another material exhibiting softening along a line in reciprocal space is ScF$_3$ \cite{ScF3}.

	In addition, it is relatively common for materials to exhibit instabilities in more than one $k$-vector \cite{Bi2SiO5,DyTe3}.
	Representative examples include the successive phase transitions in SiO$_2$ and BaTiO$_3$ \cite{SiO2,BaTiO3}, where unstable phonon modes condense one by one.
	In some cases, one mode condenses while the other remains inactive and does not participate in the phase transition \cite{DyTe3}.
	SrTiO$_3$ has been known to possess two soft modes, the antiferrodistortive mode at the R point and the ferroelectric mode at the $\Gamma$ point. 
	Both modes soften together, and the mode at the R point condenses first at 105 K \cite{STO_R-mode}.
	Although the $\Gamma$ point mode continues to soften, it does not freeze down to the absolute zero temperature due to the quantum paraelectricity.
	For slightly modified compositions, this mode also freezes.

\begin{figure}[t]
\begin{center}
\includegraphics[width=48mm]{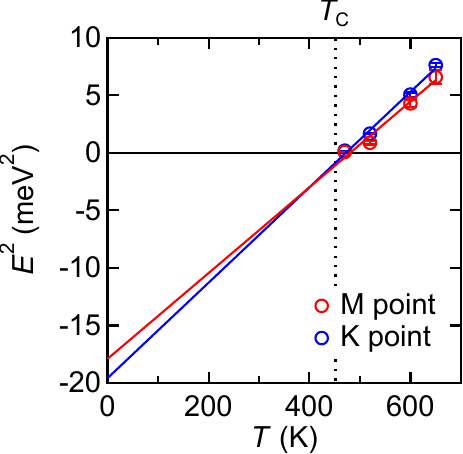}
\caption{\label{meanfield} 
Temperature dependence of the squared soft-mode energies measured at the M point (1/2 0 8) and the K point (1/3 1/3 8). The straight lines represent linear fits to the squared energies of each mode as a function of temperature.
The temperatures at which the two approximate linear fits cross $E=0$ agree within the experimental uncertainty.
This temperature is slightly higher than the reported $T_{\rm C}$ = 450 K.
Our previous single-crystal x-ray diffraction experiments show that the superlattice reflections begin to develop a precursor feature at temperatures above $T_{\rm C}$ \cite{Ishii_PRB93}, which may also be reflected in the present result.
}
\end{center}
\end{figure}

	Fig. \ref{meanfield} shows the temperature dependence of the phonon energies measured at the M point (1/2 0 8) and the K point (1/3 1/3 8).
	Both mode energies follow a mean-field-type behavior for a second-order transition, $E^2 \propto (T-T_{\rm C})$.
	Extrapolation to $T=0$ gives the energy of the K-point mode slightly smaller than that of the M-point mode.
	This result is qualitatively consistent with the DFT calculation shown in Fig. \ref{eigen}(b). 
	Condensation of the slightly more unstable K-point mode leads to a low-temperature structure that is less stable than that formed from M.
	That is, due to the small energy difference, the mode at the M point freezes and triggers the structural phase transition, whereas the mode at the K point, which is outcompeted by the M-point mode, reverts to its original energy.	
	Indeed, in the calculated phonon dispersion of the low-temperature phase, the instability at the K point disappears \cite{Ishii_PRB106, Supple}.
	Since the vibrational pattern of the K-point mode closely resembles that of the M-point mode \cite{Ishii_PRB93}, it is likely that the instability at the K point is also resolved through the condensation of the M-point mode.

	The modes at the M and K points in this system can be considered to be Rigid Unit Modes originating from the tilting of tetrahedra \cite{Perez-Mato}.
	RUMs have been known as collective vibrational modes of corner-sharing polyhedra and are widely discussed in relation to negative thermal expansion \cite{RUM1, RUM2, ZrW2O8_1, ZrW2O8_2}. 
	In addition, RUMs are often characterized as anharmonic low-energy vibrational modes.
	Representative materials exhibiting RUMs include ZrW$_2$O$_8$ and ScF$_3$.
	ScF$_3$, which shows a phonon softening along the zone boundary \cite{ScF3}, is also recognized as a structural quantum critical material, and the zone-boundary soft modes remain uncondensed due to quantum criticality. 
	Their softening does not exceed about 3 meV, while BaAl$_2$O$_4$, which lies near a structural quantum critical point, exhibits softening of both the M and K modes down to nearly $E=0$.
	While the soft mode in ScF$_3$ is an optical mode \cite{ScF3}, the present system involves the softening of an acoustic mode.
	How the phonon dynamics differ between structural quantum criticality associated with an optical mode and that associated with an acoustic mode is of great interest.

	According to the previous reports, Sr substitution suppresses the structural phase transition at the M point, resulting instead in the appearance of superlattice-like reflections at the K point \cite{Ishii_PRM}. 
	However, this behavior is not sustained, and no structural phase transition occurs at the Sr compositions higher than the structural quantum critical composition, $x=0.1$. 
	Our findings here suggest it is possible that the K-point mode could also be involved in the quantum criticality of this system. 
	The behavior of the K-point mode near the quantum critical point is of particular interest and deserves further investigation.

\section{Conclusions}
	To elucidate the mechanism of the structural phase transition of BaAl$_2$O$_4$, we have performed the inelastic x-ray scattering experiment.
	Phonon spectra collected over a temperature range from 650 K to room temperature have revealed that one of the acoustic branches softens at the M point as the temperature decreases from high temperature toward $T_{\rm C}$.
	After condensation at the $T_{\rm C}$, it generates a new acoustic branch at lower temperatures below $T_{\rm C}$.
	This observation provides experimental evidence that the ferroelectric phase transition in this system is driven by the softening of the M-point acoustic mode.
	In addition to the M-point soft mode, this system possesses another mode that softens at the K point. 
	Above $T_{\rm C}$, the K-point mode is nearly degenerate in energy with the M-point mode and softens similarly down $T_{\rm C}$.
	However, below $T_{\rm C}$ the phonons behave differently.
	The M-point mode is stabilized as an acoustic mode for the low-temperature structure while the K-point mode hardens as temperature is reduced.
	These findings offer strong evidence that structural fluctuations originating from these two dominant soft modes play a key role at the structural quantum critical point induced by Sr substitution in this system.

\begin{acknowledgements}
We thank Prof. Taniguchi (Nagoya Univ.) for helpful discussions.
This work was partially supported by JSPS KAKENHI (No. 23K17673) and an SDGs Research Project of Shimane University. 
The synchrotron radiation experiments were performed at BL35XU of SPring-8 with the approval of the Japan Synchrotron Radiation Research Institute (JASRI) (Proposal No. 2020A1179).

\end{acknowledgements}

\end{document}



\title{{\small \rm Supplemental Material} \\
\vspace{3mm}
Degenerate Soft Modes and Selective Condensation in BaAl$_2$O$_4$ via Inelastic X-ray Scattering}


\author{Yui~Ishii}
\email{yishii@mat.shimane-u.ac.jp}
\affiliation{Co-Creation Institute for Advanced Materials, Shimane University, Shimane 690-8504, Japan}

\author{Arisa~Yamamoto}
\affiliation{Department of Materials Science, Osaka Metropolitan University, Osaka 599-8531, Japan}

\author{Alfred Q. R.~Baron}
\affiliation{Materials Dynamics Laboratory, RIKEN SPring-8 Center, Sayo, Hyogo 679-5148, Japan}
\affiliation{Precision Spectroscopy Division, SPring-8/JASRI, Sayo, Hyogo 679-5198, Japan}

\author{Hiroshi~Uchiyama}
\affiliation{SPring-8/JASRI, Sayo, Hyogo 679-5198, Japan}


\author{Naoki~Sato}
\affiliation{Research Center for Materials Nanoarchitectonics (MANA), National Instititue for Materials Science (NIMS), Tsukuba, Ibaraki 305-0044, Japan}




\maketitle


	This Supplemental Material provides the following information:

\begin{itemize}
	\item Details of the modeling and fitting procedures
	\item Definitions of the lattice vectors and reciprocal lattice vectors.
	\item Phonon eigenvector analyses for the high-temperature and low-temperature phases.
	\item Phonon spectra measured at temperatures between 650 and 300$^{\circ}$C with various scattering vectors along the $\Gamma$-M and $\Gamma$-K. Several fitting parameters obtained from the fit of Figure 3.
\end{itemize}

\section{Modeling and fitting the spectra}
The measured spectra were modeled as a sum of ($N=$ 2 or 3) damped harmonic oscillator phonon lines, indexed by $j$, weighted by the usual 1-phonon detailed balance factor, $DB(x)=x/(1-e^{-x})$ (see \cite{Baron2019} and references therein). This was convolved with the resolution function, $R(E)$, as was determined by measuring plexiglass, PMMA, near its structure factor maximum, as is mostly an elastic scatterer.  The model has the form

\begin{equation}\label{dho_model}
I_{model}(E) = R(E) \otimes \left[ A_0 \delta(E) + DB\left(\frac{E}{k_BT}\right) \sum_{j=1}^{N} A_j \frac{2 \Omega_j^2 \gamma_j/\pi}{\left({E}^2-\Omega_j^2\right)^2+4{E}^2\gamma_j^2} \right]
\end{equation}
$E$ is the energy transfer, and $k_B=0.08617$ meV/K the Boltzman constant. The parameters describing each phonon line are its  energy, $\Omega_j$, width, $\gamma_j$  (the half-width at half maximum, HWHM, if $\omega_j\gg \gamma_j$) and amplitude, $A_j$.   Sometimes one considers the bare energy, $\omega_j$, given by $\omega_j^2=\Omega_j^2-\gamma_j^2$. We note the integral of each DHO term (the fraction inside the sum of eqn. \ref{dho_model}) over energy transfer is unity.

The parameters for each spectrum were optimized by minimizing the difference, $\chi^2 \equiv \sum_{i} \frac{(y_i-I_{model}(E_i))^2}{\sigma_i^2}$ (where $\sigma_i$ is the error on the $i^{th}$ data point at energy $E_i$ having measured intensity $y_i$), between the model and the data, using code based on the Minuit package \cite{James1975} as incorporated in the absfit code \cite{baron_absfit}.  The free parameters for the fitting were the amplitude, energy and width of each phonon line, $A_j$, $\Omega_j$, $\gamma_j$, and a small zero offset (typically $\lesssim$ 0.1 meV) to account for drift of the elastic energy of the spectrometer.  As the momentum, $Q$, -resolution was generally good, the dispersion of a phonon lines within a single spectrum was neglected. For most spectra, the purely elastic scattering was negligible, $A_0=0$, but even when $A_0$ was allowed to be free, originating, presumably, from parasitic scattering from the sample mounting, $A_0$ was much smaller than the phonon amplitudes.   Typical values for the reduced chi-squared, $\chi_{red}^2 \equiv \chi^2/N_{DoF}$, where $N_{DoF}$ is the number of degrees of freedom, were $\chi_{red}^2 = 0.8$ to $1.2$, indicating reasonable fits, as is also evident from the figures.

\section{Definitions of lattice vectors and reciprocal lattice vectors}

\begin{figure}[h]
\begin{center}
\includegraphics[width=55mm]{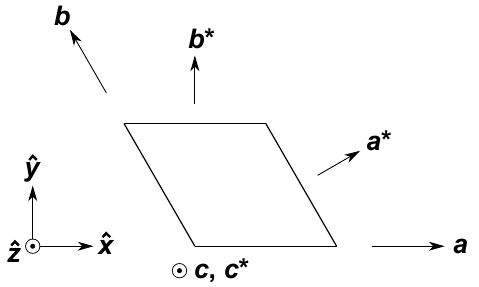}
\caption{\label{lattice} 
Relationship between the Cartesian system, lattice vectors, and reciprocal lattice vectors used in the calculation. 
}
\end{center}
\end{figure}

In the DFT calculations described in the main text, the lattice vectors are defined using a Cartesian coordinate system as follows.
\[
\bm{a} = a\hat{\bm{x}}, \; \bm{b} = -\frac{a}{2}\hat{\bm{x}} + \frac{\sqrt{3}}{2}a\hat{\bm{y}}, \; \bm{c} = c\hat{\bm{z}}
\]
The corresponding reciprocal lattice vectors are defined accordingly.
\[
\bm{a}^* = \frac{1}{a}\hat{\bm{x}} + \frac{1}{\sqrt{3} a}\hat{\bm{y}}, \; \bm{b}^* = \frac{2}{\sqrt{3} a}\hat{\bm{y}}, \; \bm{c}^* = \frac{1}{c}\hat{\bm{z}}
\]
The relationship between the Cartesian coordinate system, the lattice vectors, and the reciprocal lattice vectors is illustrated in Fig. \ref{lattice}.

\section{Analysis of phonon eigenvectors}

\begin{figure}[h]
\begin{center}
\includegraphics[width=160mm]{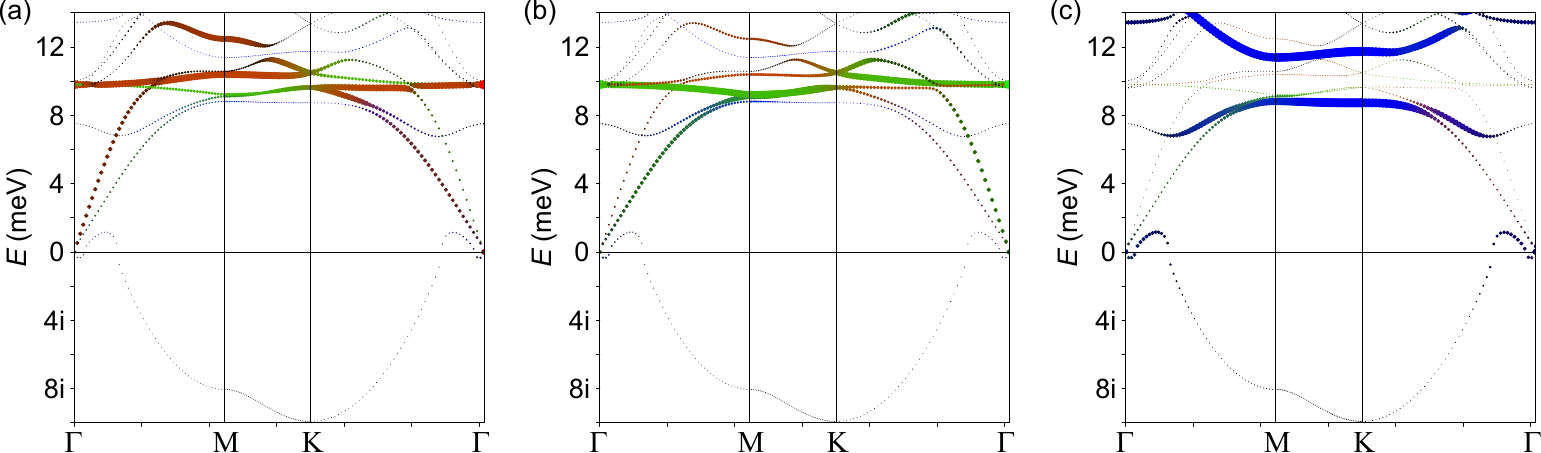}
\caption{\label{eigen_Ba} 
Analysis of phonon eigenvectors for Ba atoms. The magnitudes of (a) $x$-, (b) $y$-, and (c) $z$-components are emphasized using RGB color coding and marker size. 
Values are summed over all Ba atoms in the unit cell. 
}
\end{center}
\end{figure}

\begin{figure}[h]
\begin{center}
\includegraphics[width=160mm]{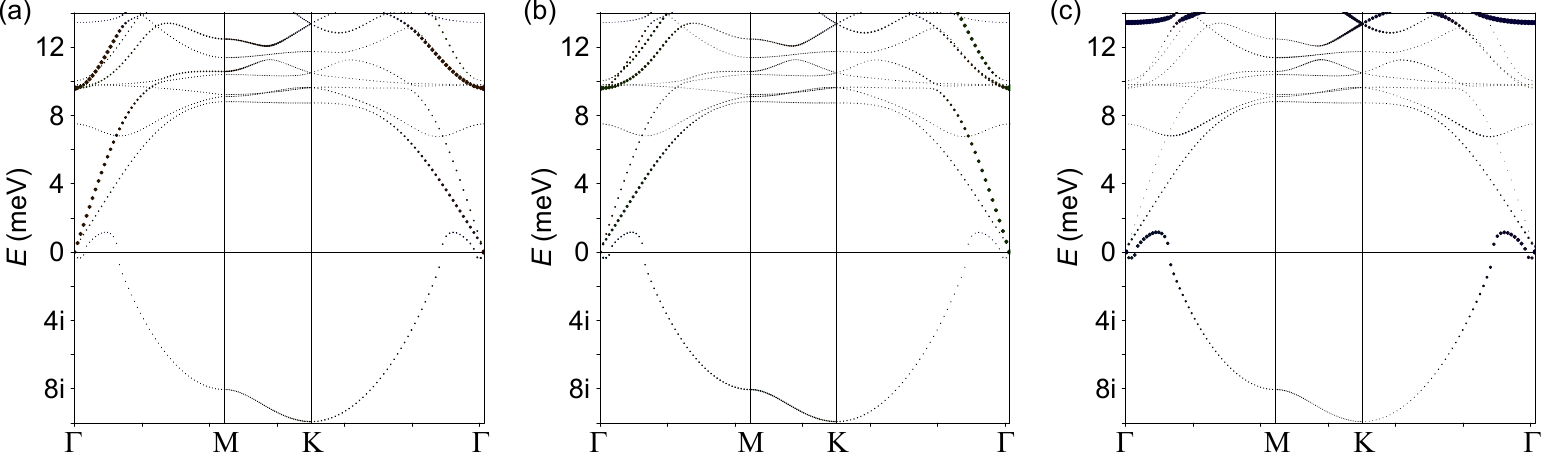}
\caption{\label{eigen_Al} 
Analysis of phonon eigenvectors for Al atoms. The magnitudes of (a) $x$-, (b) $y$-, and (c) $z$-components are emphasized using RGB color coding and marker size. Values are summed over all Al atoms in the unit cell.
}
\end{center}
\end{figure}

\begin{figure}[h]
\begin{center}
\includegraphics[width=160mm]{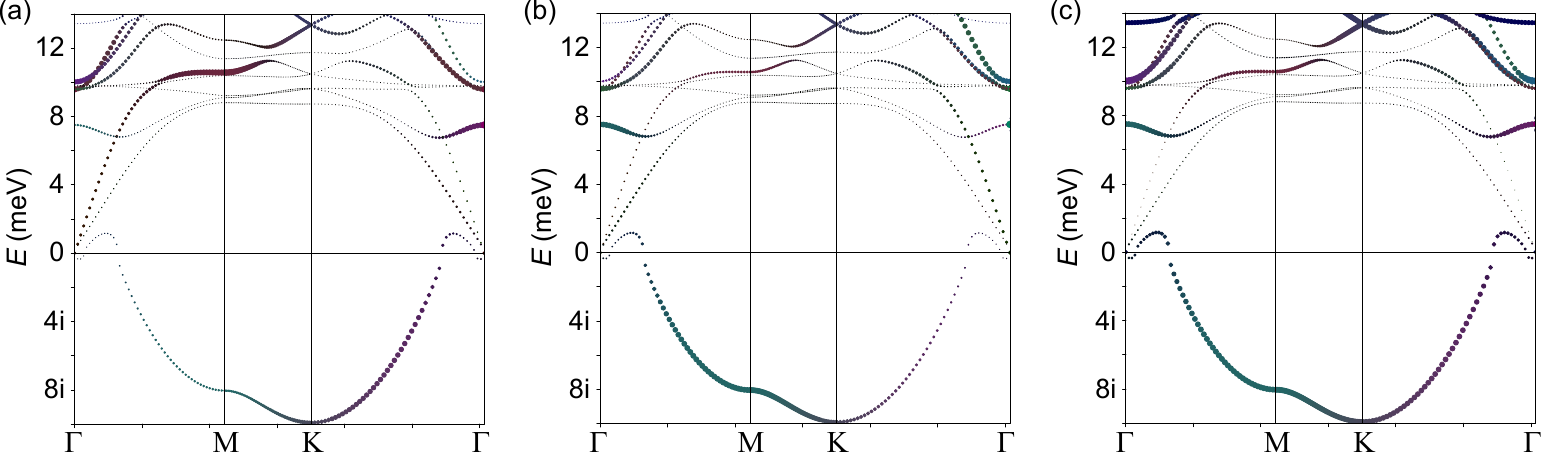}
\caption{\label{eigen_O} 
Analysis of phonon eigenvectors for O atoms. The magnitudes of (a) $x$-, (b) $y$-, and (c) $z$-components are emphasized using RGB color coding and marker size. Values are summed over all O atoms in the unit cell.
}
\end{center}
\end{figure}

	Figures \ref{eigen_Ba}, \ref{eigen_Al}, and \ref{eigen_O} represent the analyses of phonon eigenvectors of Ba, Al, and O atoms of the high-temperature (HT) phase, respectively. The eigenvectors are visualized using RGB color coding and marker size; the stronger red, green, and blue mean the larger magnitude of $x$-, $y$-, and $z$-components of phonon eigenvectors obtained for each $\bm{k}$-vector.
	
	In the HT phase of BaAl$_2$O$_4$, there are two formula units per unit cell ($Z =2$).
	Accordingly, a unit cell contains two Ba, four Al, and eight O atoms.
	The $x$-, $y$-, and $z$-components of the eigenvectors are shown in Figs. \ref{eigen_Ba}--\ref{eigen_O} as the sums over the corresponding Ba, Al, and O atoms within the unit cell.

\vspace{10mm}

\begin{figure}[h]
\begin{center}
\includegraphics[width=95mm]{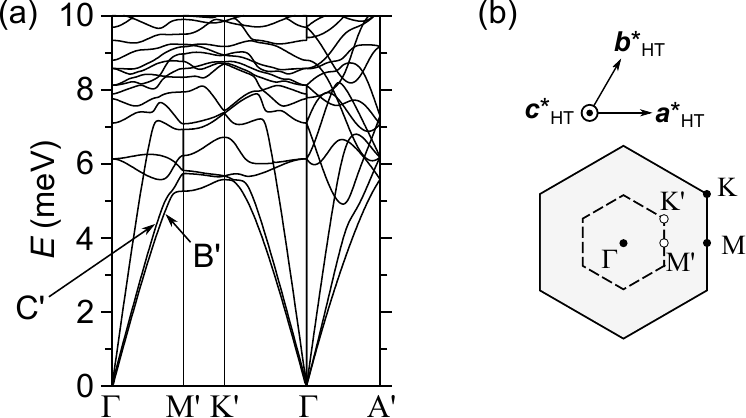}
\caption{\label{P63_disp} 
Calculated phonon dispersion relation of the low-temperature phase of BaAl$_2$O$_4$ (a space group $P6_3$). 
Supercell size of $2\times 2 \times 3$ was employed for the calculation.
(b) Relationship of BZs between the HT (solid line) and LT (broken line) phases.
}
\end{center}
\end{figure}  

\begin{figure}[h]
\begin{center}
\includegraphics[width=150mm]{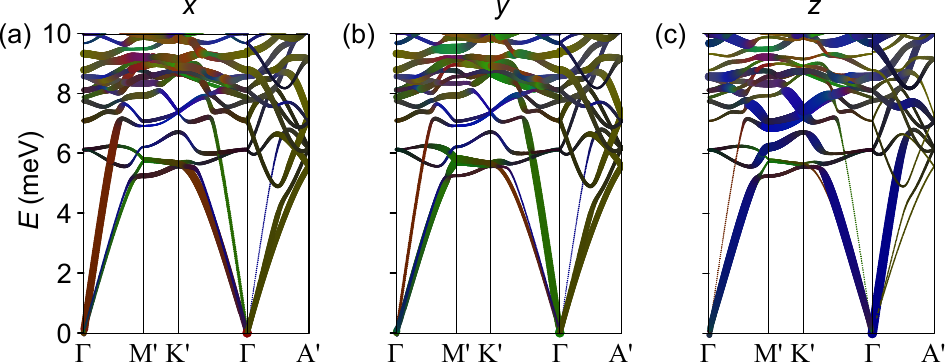}
\vspace{8mm}
\includegraphics[width=150mm]{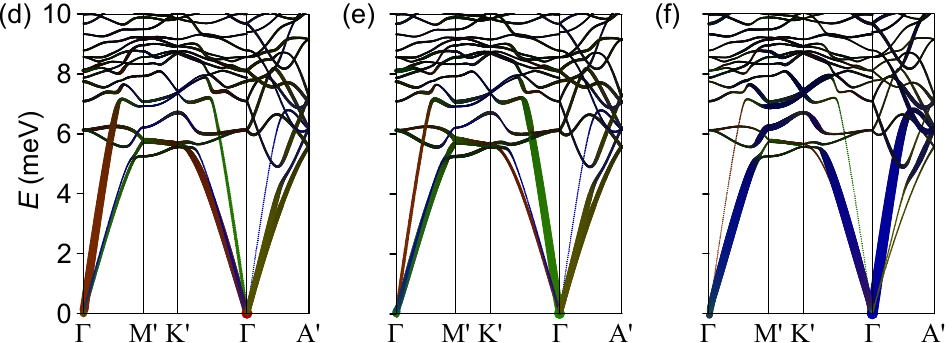}
\vspace{8mm}
\includegraphics[width=150mm]{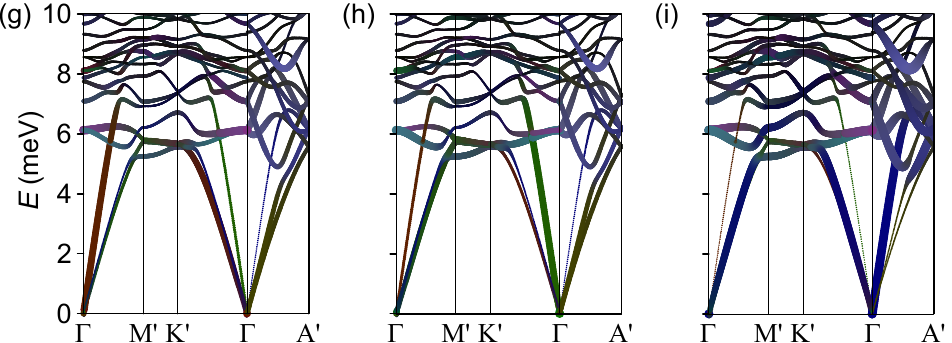}
\caption{\label{P63_eigen} 
Visualization of eigenvector components of atomic vibrations of the low-temperature phase of BaAl$_2$O$_4$. Panels (a)--(c) correspond to Ba, (d)--(f) to Al, and (g)--(i) to O atoms, with each set showing the $x$-, $y$-, and $z$-components, respectively.
}
\end{center}
\end{figure}

	We also performed phonon calculations for the low-temperature (LT) phase. 
	The space group of the LT phase is $P6_3$, and the unit cell contains eight formula units ($Z=8$).
	Compared to the HT phase, the unit-cell size is doubled along the $a$- and $b$-axes.
	The calculated phonon dispersion is shown in Fig. \ref{P63_disp}(a).
	Fig. \ref{P63_disp}(b) illustrates the relationship between the Brillouin zones of the HT and LT phases.
	Analyses of the phonon eigenvectors are presented in Figs. \ref{P63_eigen}(a)--\ref{P63_eigen}(i).
	In these figures, the same visualization method as in Figs. \ref{eigen_Ba}--\ref{eigen_O} is applied, where color coding and marker size represent the magnitude of each atomic eigenvector.
	The left, center, and right columns correspond to the $x$-, $y$-, and $z$-components, respectively.
	Panels (a)--(c) show the results for Ba atoms, (d)--(f) for Al atoms, and (g)--(i) for O atoms.

\clearpage
\section{Temperature-dependent phonon spectra measured at various $Q$ vectors}

\begin{figure}[h]
\begin{center}
\includegraphics[width=160mm]{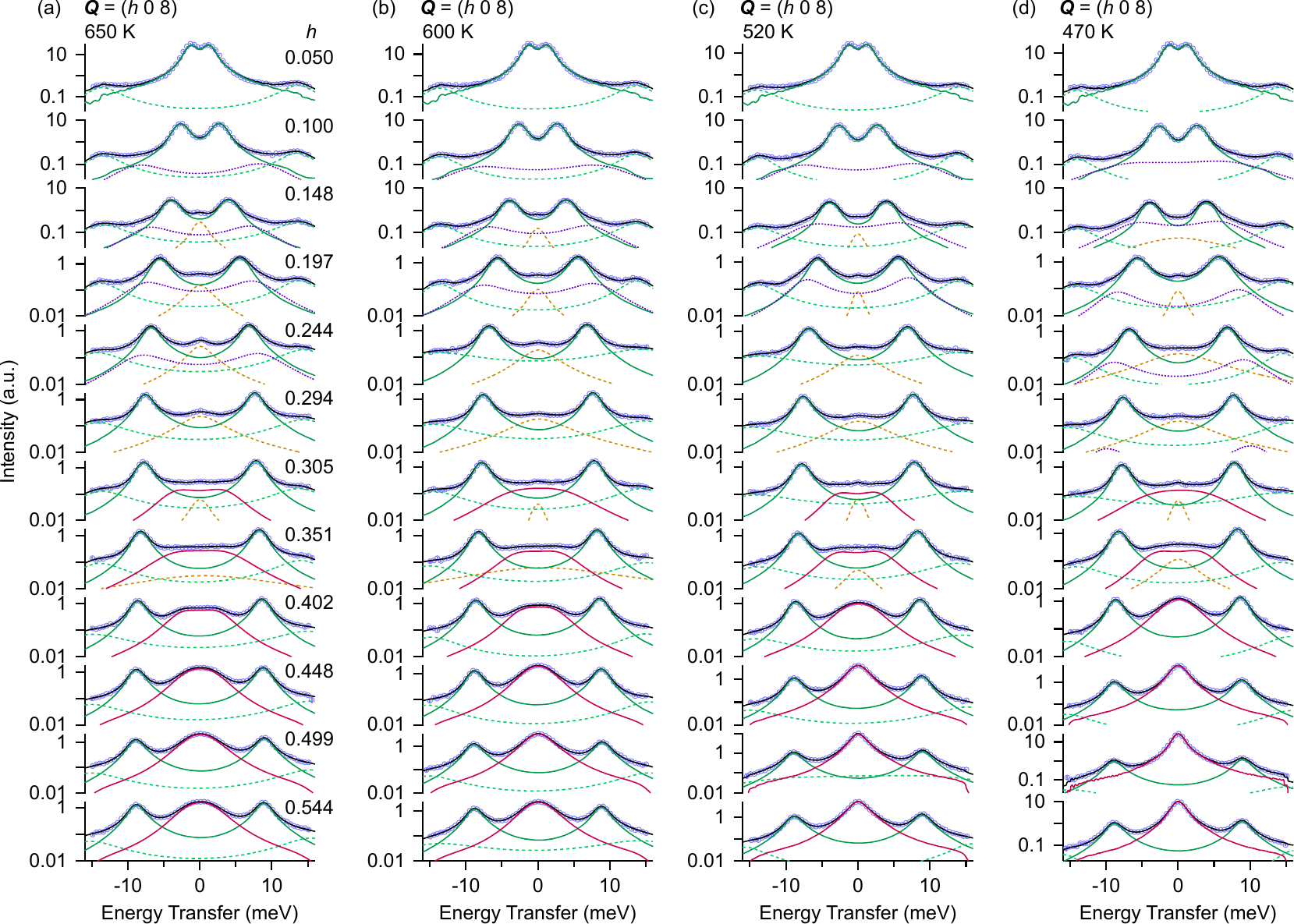}
\caption{\label{spectra_GMHT} 
IXS spectra measured at  (a) 650, (b) 600, (c) 520, and (d) 470 K along the $\Gamma$-M direction with scattering vectors $\bm{Q} = (h \; 0 \; 8)$. Although the fitting result shown by the red curve at $h = 0.305$ is of low reliability, its inclusion was essential for achieving a reasonable fit.
}
\end{center}
\end{figure}

\begin{figure}[h]
\begin{center}
\includegraphics[width=160mm]{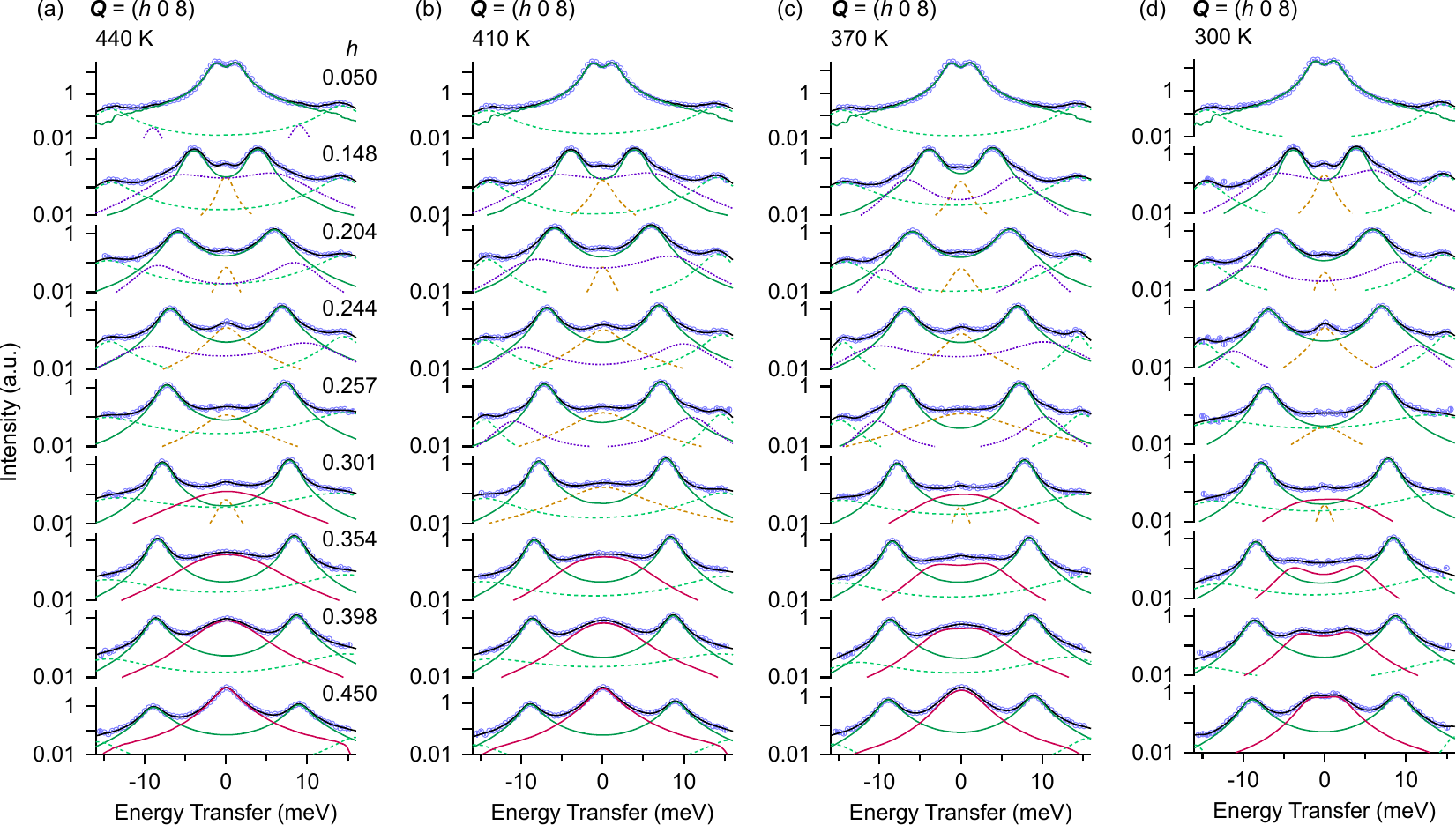}
\caption{\label{spectra_GMLT} 
IXS spectra measured at  (a) 440, (b) 410, (c) 370, and (d) 300 K along the $\Gamma$-M direction with scattering vectors $\bm{Q} = (h \; 0 \; 8)$. Although the fitting result shown by the red curve at $h = 0.301$ is of low reliability, its inclusion was essential for achieving a reasonable fit.
}
\end{center}
\end{figure}

\begin{figure}[h]
\begin{center}
\includegraphics[width=160mm]{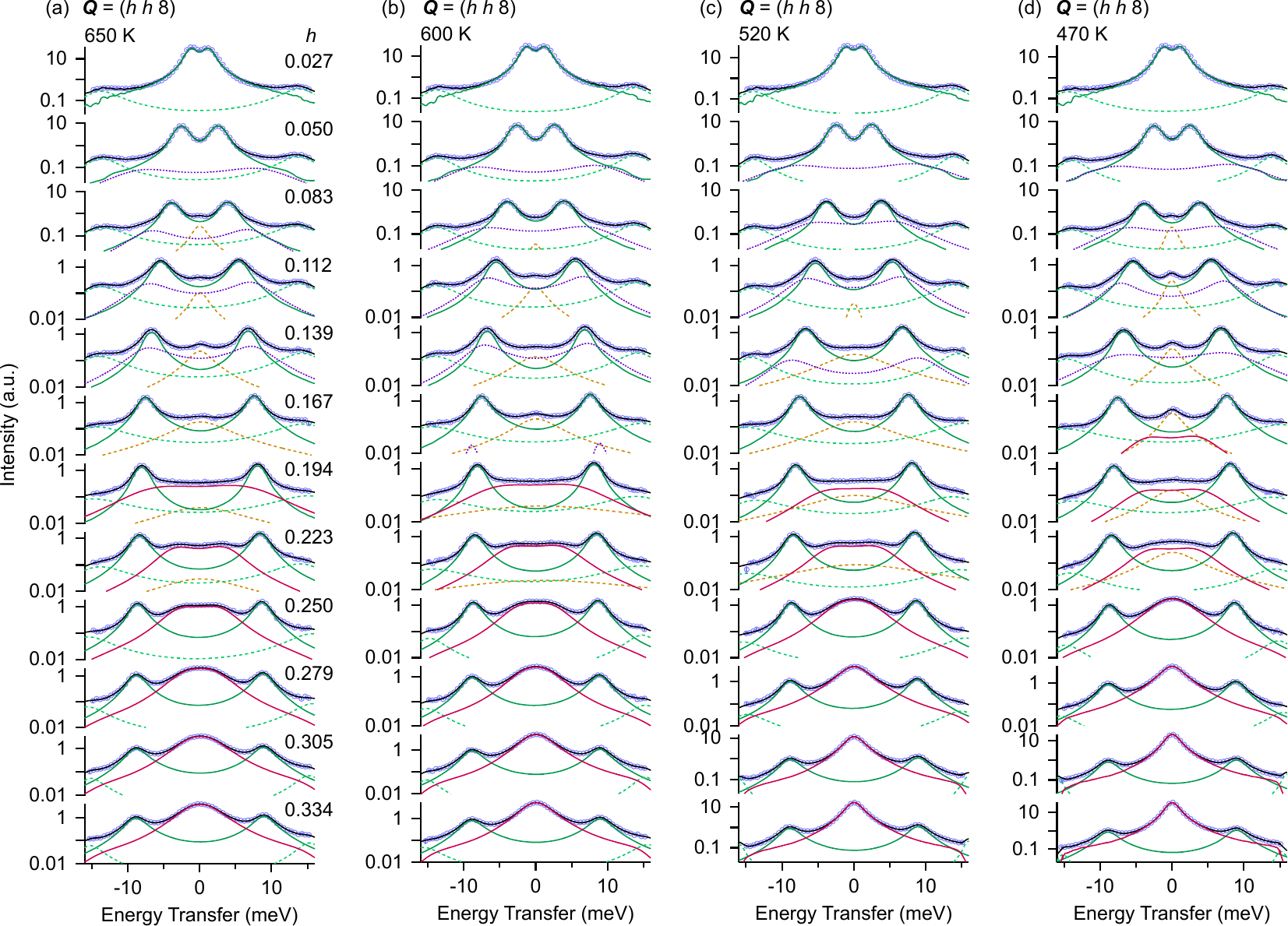}
\caption{\label{spectra_GKHT} 
IXS spectra measured at  (a) 650, (b) 600, (c) 520, and (d) 470 K along the $\Gamma$-K direction with scattering vectors $\bm{Q} = (h \; h \; 8)$. Although the fitting result shown by the red curve at $h = 0.194$ is of low reliability, its inclusion was essential for achieving a reasonable fit.
}
\end{center}
\end{figure}

\begin{figure}[h]
\begin{center}
\includegraphics[width=160mm]{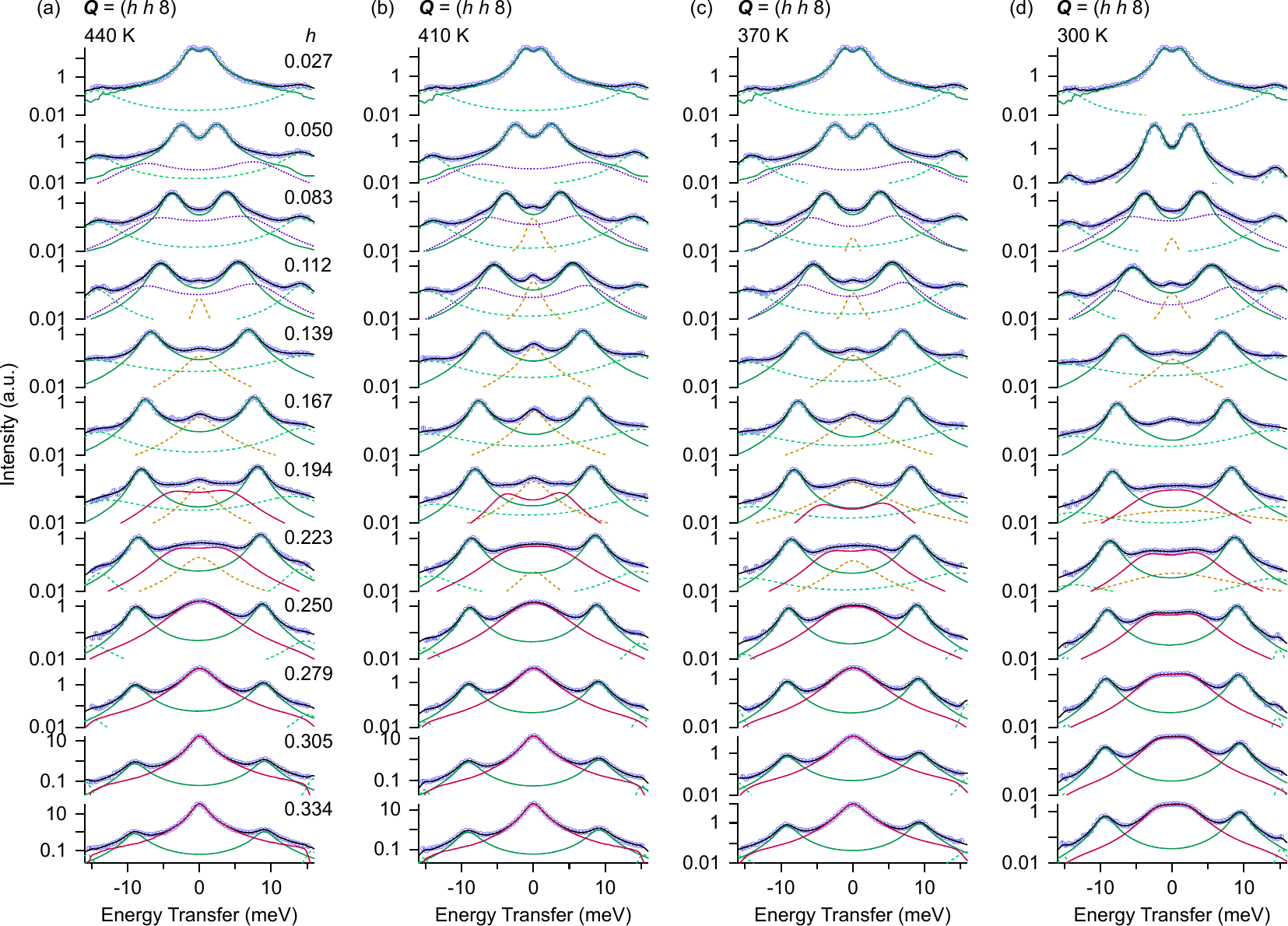}
\caption{\label{spectra_GKLT} 
IXS spectra measured at  (a) 440, (b) 410, (c) 370, and (d) 300 K along the $\Gamma$-K direction with scattering vectors $\bm{Q} = (h \; h \;8)$. Although the fitting result shown by the red curve at $h = 0.194$ is of low reliability, its inclusion was essential for achieving a reasonable fit.
}
\end{center}
\end{figure}

	This section provides all phonon spectra measured in this study. Figs. \ref{spectra_GMHT} and \ref{spectra_GMLT} correspond to the results for the $\Gamma$-M measurement. Figs. \ref{spectra_GKHT} and \ref{spectra_GKLT} correspond to the results for the $\Gamma$-K measurement.   
	Each spectrum is successfully resolved into several peaks using the DHO model. 
	A small elastic peak at $E=0$ is included when necessary, as shown by broken orange curves.
	Solid green, dotted purple, and broken green curves represent the contributions from branches B, D, and E, respectively.
	Solid red curves represent the contribution from the soft modes.
	
\vspace{3mm}

\begin{figure}[h]
\begin{center}
\includegraphics[width=120mm]{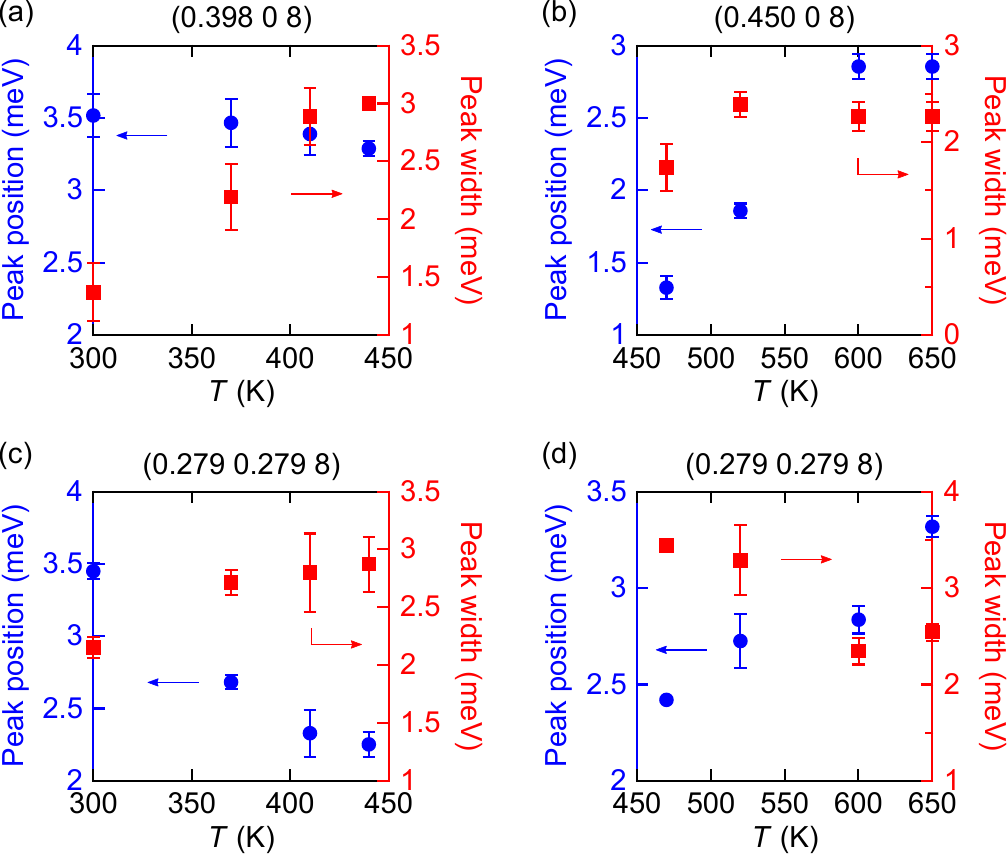}
\caption{\label{width} 
The fitting parameters, peak position and peak width, for the soft modes M and K obtained from the fit of Figure 3 in the main text. (a) and (b) correspond to the parameters of the M-point mode below and above the $T_{\rm C}$, respectively. (c) and (d) correspond to the K-point mode.
}
\end{center}
\end{figure}